\documentclass[]{article}
\usepackage{emulateapj}
\usepackage{psfig}
\begin{document}

\title{QUASI-PERIODIC VARIABILITY AND THE INNER RADII OF THIN
  ACCRETION DISKS IN GALACTIC BLACK-HOLE SYSTEMS}

 \author{Tiziana Di Matteo\altaffilmark{1} and Dimitrios Psaltis}
 \affil{Harvard-Smithsonian Center for Astrophysics,
60 Garden St., Cambridge, MA 02138;\\
tdimatteo, dpsaltis@cfa.harvard.edu}

\altaffiltext{1}{{\em Chandra\/} Fellow}

 \begin{abstract} We calculate upper bounds on the inner radii of
   geometrically thin accretion disks in galactic black-hole systems
   by relating their rapid variability properties to those of neutron
   stars. We infer that the inner disk radii do not exhibit large
   excursions between different spectral states, in contrast with the
   concept that the disk retreats significantly during the
   soft-to-hard state transition.  We find that, in the hard state,
   the accretion disks extend down to radii $\lesssim 6-25\;GM/c^2$
   and discuss the implications of our results for models of
   black-hole X-ray spectra.  \end{abstract}


\section{INTRODUCTION}

X-ray spectra and rapid variability provide discriminating signatures
for the properties of accreting compact objects. Correlating these
characteristics in the case of weakly-magnetic neutron stars has
provided strong constraints on physical models for their accretion
flows (see van der Klis\markcite{vdk98} 1998 for a review). The
detailed timing studies of galactic X-ray sources with the {\em Rossi
X-ray Timing Explorer\/} ({\em RXTE\/}) make now possible the
extension of similar analyses to the case of accreting black holes.

Accreting black holes show a variety of X-ray spectral states (Tanaka
\& Lewin\markcite{TL95} 1995). The majority of sources are usually
observed in the so-called hard/low and soft/high states named after
the relative hardness of the X-ray spectra and the luminosity in the
soft ($\sim 0.5-2$~keV) X-ray band. In the standard paradigm, the soft
and hard X-ray spectral components are attributed, respectively, to a
geometrically thin accretion disk and a hot Comptonizing medium, with
their relative contribution determining the source state (as is the
case for coronal models [see, e.g., Haardt \& Maraschi\markcite{HM93}
1993; Zdziarski et al.\markcite{Zetal98} 1998; Poutanen \&
Coppi\markcite{PC98} 1998; Dove et al.\markcite{Detal97} 1997 for
recent studies], advection-dominated accretion flows [e.g., Esin et
al.\markcite{Eetal97} 1997], etc.). In most cases, the X-ray spectra
of the hard/low state show no evidence for soft blackbody emission
(e.g., Esin et al.\ 1997) or strong reflection features (i.e., the
backscattered emission from the putative accretion disk; see, e.g.,
Gierli\'nski et al.\markcite{Getal97} 1997). Based on this, it has
been argued that, in the hard state, the geometrically thin disks do
not extend down to the innermost stable orbit but are truncated at
tens or hundreds of Schwarzchild radii.

The rapid variability properties of black holes can be used in
assessing the accretion-flow geometries inferred from spectral models.
Various types of quasi-periodic oscillations (QPOs) and peaked noise
features have been observed from many persistent and transient
black-hole sources (e.g., van der Klis\markcite{vdk95} 1995). The
frequencies of these variability components follow correlations that
are consistent with the ones observed in neutron-star systems
(Wijnands \& van der Klis\markcite{WK99} 1999; Psaltis, Belloni, \&
van der Klis\markcite{PBK99a} 1999a). They are, therefore, likely to
be produced by similar physical mechanisms. In particular, the QPO
properties in both black-hole and neutron-star systems vary
systematically and reproducibly with spectral state and can have
fractional widths down to $\delta \nu/\nu \lesssim 0.2$ (see van der
Klis 1995). For this reason, although they are related to the hard ($>
2$~keV) component of the X-ray spectrum, their frequencies are
believed to be determined by modulations in the geometrically thin
component of the accretion flow.

In this {\em Letter}, we use the empirical correlations between QPO
frequencies in neutron-star and black-hole systems to infer the
fastest variability timescales for specific black-hole spectral
states. This information allows us to place upper bounds on the inner
radii of the geometrically thin components of the accretion flows and
compare them to those inferred by the spectral models of galactic
black holes. Even though we use the empirically determined
correlations, our results are supported by the recent identification
of the various observed QPO frequencies with fundamental
general-relativistic frequencies around compact objects (Stella \&
Vietri\markcite{SV99} 1999; Stella, Vietri, \& Morsink\markcite{SVM99}
1999; Psaltis \& Norman\markcite{PN99} 1999).

\section{SPECTRA AND RAPID VARIABILITY OF ACCRETING BLACK HOLES}

We use previously published data for a sample of four persistent and
six transient galactic black-hole systems. These include the best
studied sources for which QPOs have been detected and the
corresponding spectral states have been identified and reported in the
literature.  Following Wijnands \& van der Klis (1999) and Psaltis et
al.\ (1999a), we include the QPOs observed in the microquasars
GRS~1915$+$105 and GRO~J1655$-$40 only when these sources show canonical
black-hole spectral states.

{\em Cyg~X-1.---\/}This persistent source spends most of its time in
the hard state but occasionally shows a transition to the soft state
(Cui et al.\markcite{Cetal97} 1997 and references therein). We use the
results of the joined {\em GINGA\/}/{\em OSSE\/} observation of
Cyg~X-1 during its hard state, in which $\simeq 1-3$~Hz QPOs were
observed (Rutledge et al.\markcite{Retal99} 1999 and references
therein). In a recent transition to the soft state, $\simeq 4-12$~Hz
QPOs were also detected with {\rm RXTE}, but no spectral analysis has
been reported for this observation (Cui et al.\ 1997).

{\em GX~339$-$4.---\/}This is one of the few sources in which five
distinct spectral states have been observed (Wilms et
al.\markcite{Wetal99} 1999 and references therein). A $\simeq
0.3-0.4$~Hz QPO has been recently detected during the hard state with
{\em RXTE\/} (Nowak, Wilms, \& Dove\markcite{Netal99} 1999) as well as
a $\simeq 6-7$~Hz during the very high state with {\em GINGA\/}
(Miyamoto et al.\markcite{Metal91} 1991).

{\em 1E~1740.7$-$2942 and GRS~1758$-$258.---\/}These two sources have
only been observed in their hard states. Recent {\em RXTE\/}
observations have revealed the presence of a 2.0~Hz and a 0.4~Hz QPO
in 1E~1740.7$-$2942 and GRS~1758$-$258, respectively (Smith et
al.\markcite{Setal97} 1997). No detailed spectral analysis is
available for these observations, but a canonical hard-state photon
index of $\simeq 1.6$ was reported for both sources by Smith et al.\ 
(1997).

{\em GRO~J1655$-$40.---\/}During the decay phase of the recent
outburst of this superluminal source, four canonical black-hole
spectral states were identified (M\'endez, Belloni, \& van der
Klis\markcite{Metal98} 1998a). In the hard and intermediate states, a
$\simeq 0.8$~Hz and a $\simeq 6.5$~Hz QPO were detected, respectively.

{\em GS~1124$-$683 (Nova Muscae 1991).---\/}During the 1991 outburst
of this source, five distinct spectral states were observed with
{\em GINGA\/} and {\em OSSE\/} (see, e.g., van der Klis 1995 and
references therein). When the source was probably in the very high
state a $\simeq 4-6$~Hz QPO was detected (Miyamoto et al.\ 1994;
Belloni et al.\ 1997).

{\em 4U~1630$-$47.---\/}This recurrent transient has been recently
observed again in outburst with {\em RXTE}. While the source was in
the hard state, a $\sim 0.8$~Hz QPO was detected (McCollough et al.\ 
1999).

{\em GRO~J0422$+$32.---\/}{\em OSSE\/} and {\em ASCA\/} have observed
this transient source in both the hard and soft states.  A $\sim
0.23$~Hz QPO was detected during the {\em OSSE\/} observation, while
the source was in the hard state (Grove et
al.\markcite{Getal98a}\markcite{Getal98b} 1998a, 1998b).

{\em XTE~J1555$-$324.---\/}{\em RXTE\/} observations of this X-ray
nova revealed characteristic spectral states of black-hole sources
(Revnivtsev, Gilfanov, \& Churazov\markcite{RGC99} 1999). During the
decay of the outburst, a $\simeq 2-3$~Hz QPO was detected.

{\em XTE~J1550$-$564.---\/}This newly discovered X-ray transient shows
QPOs with frequencies that vary in the range $\sim 0.1-10$~Hz as its
energy spectrum softens (Cui et al.\markcite{Cetal1998} 1998). Here we
use only the data points with QPO frequencies $\gtrsim 1$~Hz, for
which the energy spectral indices were reported in the literature
(Sobczak et al.\markcite{Setal99} 1999).

Figure~1 shows the photon spectral indices and the corresponding  
QPO frequencies for the sources in our sample.

\section{THE INNER RADII OF THE ACCRETION DISKS}

The observed QPOs in accreting compact objects are typically narrow
($\delta\nu/\nu\lesssim 0.2$) with properties that vary systematically
and reproducibly with spectral state. Although QPOs are associated
with the hard X-ray spectral components (see, e.g., Berger et al.\
1996), the above characteristics can be accounted for in a physical
model only if the timing properties of the QPOs are associated with
the geometrically thin component of the accretion flow. Indeed, for
all frequencies in the flow, $\delta\nu/\nu\sim (H/R)^2$, and hence
$H/R\ll 1$. This argument is model-independent and has been the basis
of most theoretical models for neutron-star and black-hole QPOs (see
Psaltis et al.\ 1999b for a recent review of QPO models). In such
models, Comptonization in a hot medium that surrounds the accretion
disk typically amplifies the oscillations (Lee \&
Miller\markcite{LM98} 1998; see also Miller et al.\ 1998), possibly
accounting for the fact that strong QPOs are observed only when the
hot component of the accretion flow is present.

\vbox{ \centerline{
    \psfig{file=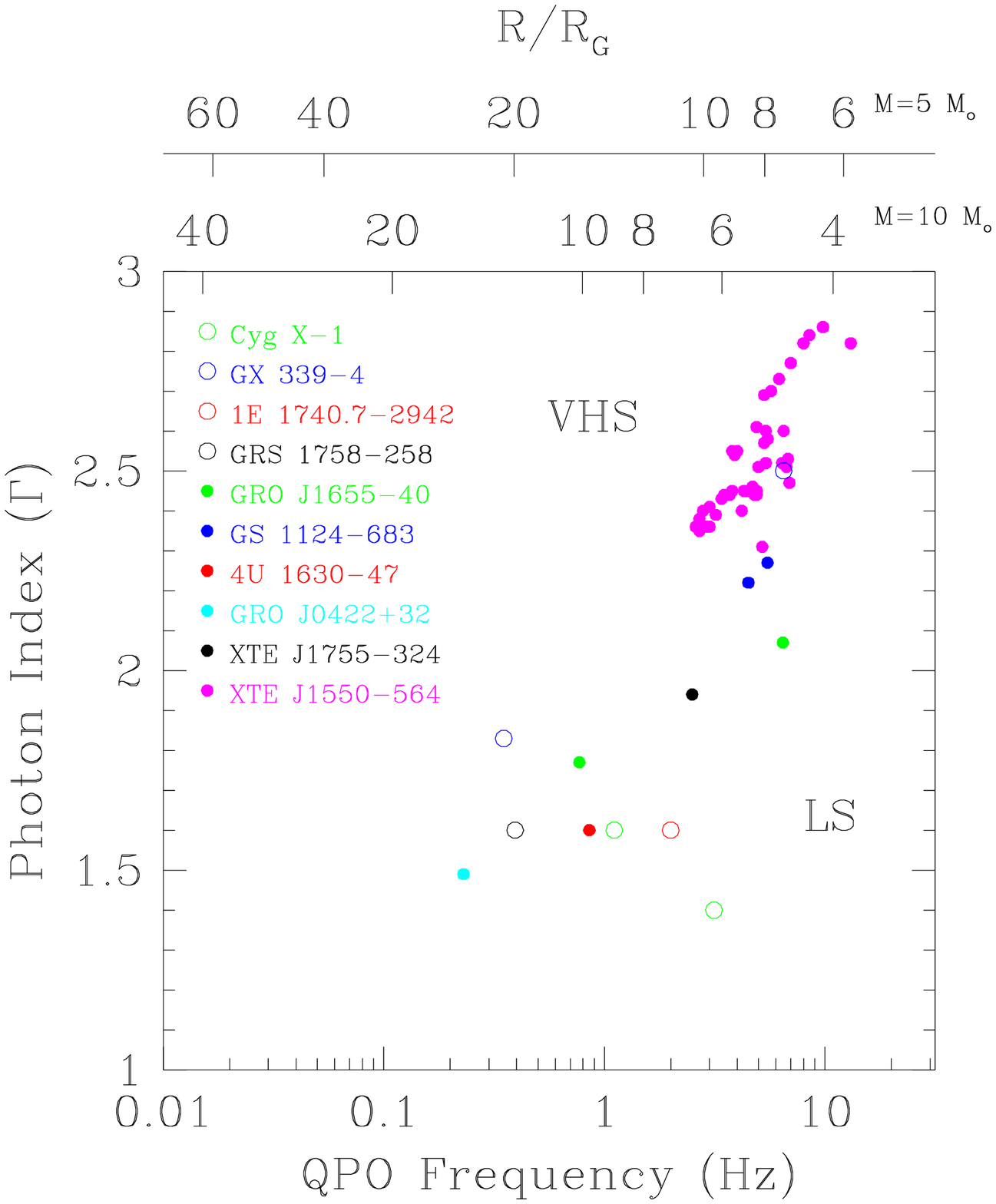,width=9.0truecm} } 
    \figcaption[]{
    \footnotesize Photon spectral index versus QPO frequency $\nu_1$ for
    different spectral states (LS: hard/low state; VHS: very high
    state) of galactic black-hole sources. The top axis shows the
    inferred upper bounds on the inner radii of geometrically thin
    accretion disks for different black hole masses.}
}
\vspace*{0.5cm}

The centroid frequency $\nu_{\rm QPO}$ of the QPO provides an
upper bound on the inner radius $R_{\rm in}$ of the geometrically thin
component of the accretion flow. Since the fastest variability
timescale at any radius around a compact object is the Keplerian
orbital frequency, then
 \begin{eqnarray}
 \nu_{\rm QPO}&\le& \frac{1}{2\pi}
   \left(\frac{GM}{R_{\rm in}^3}\right)^{1/2}\nonumber
 \end{eqnarray}
and hence
 \begin{equation}
 \left(\frac{R_{\rm in}}{R_{\rm g}}\right) \le 220\; \nu_{\rm QPO}^{-2/3} 
   \left(\frac{M}{10~M_\odot}\right)^{-2/3}\;,
 \label{nu_k}
 \end{equation}
 where $R_{\rm g} \equiv GM/c^2$, and $M$ is the mass of the compact
 object. In this section we deduce the fastest variability
 timescales during different black-hole spectral states and use
 relation~(\ref{nu_k}) to place upper bounds on the inner radii of the
 accretion disks.

 The power-density spectra of accreting black holes typically show a
 broad noise component that is flat-topped up to a break frequency
 $\nu_{\rm b}$ and decreases as a power law above it. Often, a narrow
 QPO at a frequency $\nu_1$ and a broader peaked noise component at a
 frequency $\nu_2$ are also detected (typically $\nu_2\sim 10 \nu_1
 \sim 10^2 \nu_{\rm b}$; see, van der Klis 1995; Psaltis et al.\ 
 1999a). When present, the QPO at $\nu_1$ is typically narrow ($\delta
 \nu/\nu\lesssim 0.2$) and unambiguously detected; this is the QPO
 we discuss in \S2 for the sources in our sample.
 
In accreting, weakly-magnetic neutron stars, the same variability
components are observed in the power spectra, together with a third
QPO peak at a frequency $\nu_3 > \nu_2$ (van der Klis 1998; Psaltis et
al.\ 1999a). In such systems, the QPOs with frequencies $\nu_1$,
$\nu_2$, and $\nu_3$ are often called the HBO, lower kHz QPO, and
upper kHz QPO, respectively (see van der Klis 1998).  The frequencies
of all these power-spectral components vary over wide ranges, often by
a factor of $10^2$ in a given source, and yet $\nu_1$ follows simple
power-law relations ($\chi^2$/d.o.f.$\sim 1$; see Psaltis et al.\
1998, 1999a for a thorough statistical analysis) with $\nu_{\rm b}$,
$\nu_2$, and $\nu_3$ (Wijnands \& van der Klis 1999; Psaltis et al.\
1999a). These correlations depend only weakly (if at all) on the other
properties of the sources thus demonstrating that they are produced in
the accretion flows and not by the compact objects.

In neutron-star systems, the fastest observed timescale is given by
the QPO at frequency $\nu_3$, which follows in all sources the
empirical correlation $\nu_1\simeq 63(\nu_3/1$~kHz$)^{1.9}$~Hz
(Psaltis et al.\ 1999a, 1999b). This correlation together with
relation~(\ref{nu_k}) gives an upper bound on the inner
radius of the accretion disk, i.e.,
 \begin{equation}
 \left(\frac{R_{\rm in}}{R_{\rm g}}\right)\le 27\;\nu_1^{-0.35}
    \left(\frac{M}{2~M_\odot}\right)^{-2/3}\;.
 \label{rin}
 \end{equation}
Observation of a narrow $\nu_1=70$~Hz QPO in a neutron-star system,
for example,  constrains the inner radius of the
geometrically thin accretion flow to be comparable to the radius of
the innermost stable orbit.

In black-hole systems, the frequencies $\nu_{\rm b}$, $\nu_1$, and
$\nu_2$ also vary widely, but follow the same correlations (with
systematic differences in the normalization of less than a factor of
two) as those exhibited by neutron-star sources (Psaltis et al.\
1999a). This implies that the three variability components with
frequencies $\nu_{\rm b}$, $\nu_1$, and $\nu_2$ are each produced by
similar physical mechanisms in both types of compact objects. As a
result, neither $\nu_1$ nor $\nu_2$ are likely to be associated with
the highest Keplerian frequencies in the accretion disks and hence
cannot be used in constraining the inner extent of the accretion
disks. Moreover, QPOs at $\nu_3$, which could provide this constraint
directly, have not been detected from any black-hole
system\footnote{It is, unclear what determines the amplitudes and
detectability of these QPOs in neutron-star systems and hence the
reason for their absence from black-hole power spectra (see, Psaltis
et al.\ 1999a for a discussion).  Indeed, there is an example of a
transient neutron-star source (4U~1608$-$52; M\'endez et al.\ 1998b)
in which the QPO at $\nu_2$ is strong and can be directly detected,
but the QPO at $\nu_3$ is revealed only after the data are analyzed
using the so-called shift-and-add technique.}.  We conjecture that the
same physical mechanism in both neutron stars and black holes produces
QPOs at $\nu_1$ and, therefore, even in black holes, $\nu_1$ is
related to the inner radius of the accretion disk as in
equation~(\ref{rin}).

Physical motivation for this hypothesis is provided by recent
modeling of the correlations between the QPO frequencies in neutron-star
and black-hole systems (Stella \& Vietri 1999; Stella et al.\ 1999; see
also Psaltis \& Norman\markcite{PN99} 1999). In these models, $\nu_3$,
$\nu_2$, and $\nu_1$ are identified, respectively, with the orbital,
periastron precession, and twice the nodal precession
general-relativistic frequencies of a perturbed orbit around the compact
object. Given an observation of a nodal precession frequency in a source,
the radius of the orbit, and hence an upper bound on the inner radius of
the accretion disk, is (see Stella et al.\ 1999) \begin{equation}
 \frac{R_{\rm in}}{M} \le \left( 
 \frac{2\alpha}{\pi}\right)^{1/3} \nu_1^{-1/3} M^{-2/3}\;, \label{GR}
 \end{equation}
 where $\alpha$ is the specific angular momentum of the
compact object and we have set $G=c=1$. The scaling in
equation~(\ref{GR}) is similar to the empirically inferred one, giving
physical motivation to relation~(\ref{rin}).
 
Relation~(\ref{rin}) provides a bound on the inner radius of the
geometrically thin accretion disk for each source in our sample.  The
inferred bounds are shown in Figure~1, for two different values of the
black-hole mass. In all spectral states in which narrow QPOs have been
detected, the accretion disks appear to extend down to $\lesssim
6-25\;R_{\rm g}$. Note here, that the inferred bounds are consistent
with the inner radii of accretion disks in neutron-star sources, as
determined by the frequencies of the kHz QPOs (see, e.g., Miller,
Lamb, \& Psaltis\markcite{MLP98} 1998).

\section{DISCUSSION}
 
The upper bounds on the inner accretion disk radii derived in \S3
imply that, in general, the geometrically thin disk component extends
close to the innermost stable orbit, independently of the specific
spectral state which the accreting black hole displays.  Our findings
are complementary to the constraints on the geometry of the accretion
flows around galactic black holes derived on the basis of radiation
physics and spectral analysis of the different spectral states
(Poutanen, Krolik \& Ryde\markcite{PKR97} 1997; Gierli\'nsky et al.\
1997; Poutanen \& Coppi 1998; Zdziarski et al.\ 1998; Done \& \.Zycki
1998; Esin et al.\ 1997; Wilms et al.\ 1999 and references
therein). In particular, in such models it is argued that, because the
reflection features in the hard state of black-hole systems are much
less prominent (e.g., the amplitude of Compton reflection
$\Omega/(2\pi)$ is $\sim 0.2-0.3$ and its associated Fe K$\alpha$
lines are weak) than in Seyfert galaxies, the inner radius of the disk
cannot be very close to the black hole, in order to subtend a smaller
solid angle to the X-ray emitting region (e.g., Gierli\'nski et al.\
1997).

In these spectral models, it is often postulated that, inside the
truncated disk, a hot central medium/corona--like structure produces
the observed hard component of the spectrum. The transition to soft
state is then assumed to occur as the geometrically thin disk moves
inwards and the dissipated energy emerges in the form of a
blackbody-like spectrum (see, however, Di Matteo, Celotti, \&
Fabian\markcite{DCF99} 1999). An alternative possibility, which
includes a dynamical description of the Comptonizing medium, has been
discussed within the context of advection-dominated accretion flows
(ADAF; Esin, McClintock, \& Narayan 1997); in these models, the inner
Compton medium in the hard state is identified with an advection
dominated region, outside of which the thin disk exists. 

Figure~1 shows that the QPO frequencies observed in the hard and
very-high states of all sources in our sample lie in a rather
restricted range. Even though the presence of a transition in the
accretion disk properties cannot be excluded, Figure~1 implies that
the inner radii of the accretion disks cannot show large excursions
between the different spectral states. Therefore, the concept of a
geometrically thin disk retreating significantly during the
soft-to-hard state transition is not supported by our results.  (Note
though that for some sources there seems to be weak evidence for a
systematic change in the QPO frequency with spectral state.)  In
agreement with the presence of a soft X-ray excess in the observed
spectra of black-hole systems in the very-high state, we find that the
inferred inner accretion disk radii are comparable to the radius of
the innermost stable orbit. On the other hand, for all sources in our
sample, the inner accretion disk radii in the hard state are inferred
to be {\em only\/} $\sim 6-25\;R_{\rm g}$ (see Fig.~1).

To illustrate this point explicitly, we discuss specific sources for
which detailed spectral modeling has been performed and compare the
inferred inner disk radii to the bounds deduced here based on their
rapid variability. For the hard state of Cyg~X-1, the truncation
radius of the accretion disk is inferred to be $\sim 30-50\;R_{\rm g}$
in coronal models (Poutanen et al.\ 1997; Poutanen \& Coppi 1998; Done
\& \.Zycki 1999) and $\sim 200 \;R_{\rm g}$ (or strictly $>60 R_{\rm
  g}$) in ADAF models (Esin et al.\ 1998). On the other hand, the
detection of a $\gtrsim 1$~Hz QPO in the same state implies $R_{\rm
  in} < 8-12\;R_{\rm g}$, depending on the mass of the black hole. For
the hard state of GX~339$-$4, $R_{\rm in} \gtrsim 150\;R_{\rm g}$ for
both coronal and ADAF models (Zdziarski et al.\ 1998; Wilms et al.\ 
1999), whereas the rapid variability constraints imply $R_{\rm in} <
12-20\;R_{\rm g}$.  In GRO~J0422$+$32, ADAF models require $R_{\rm in}
\sim 10^4\;R_{\rm g}$ (Esin et al.\ 1997), while the presence of a
$\simeq 0.23$~Hz QPO constraints the inner disk radius to be $R_{\rm
  in} < 15-25\;R_{\rm g}$.

In all the above sources, the discrepancy between the inner disk radii
inferred from the rapid variability and those reported from spectral
studies can be as large as a few orders of magnitude. However, in many
spectral models, $R_{\rm in}$ was not necessarily chosen to be the
smallest value consistent with the observed X-ray spectra. Here we
showed that rapid variability properties impose upper bounds on the
inner extent of the geometrically thin accretion disks in galactic
black-hole systems. We, therefore, argue that combining both the
variability and spectral properties of accreting black holes provides
complementary constraints on models of their inner accretion
flows. Our constraints on the inner radii will become unambiguous with
the detection of quasi-periodic variability in the hard state of black
hole systems with frequencies $\nu_3 \gtrsim 100$~Hz. Such a detection
would also impose a strict upper bound on the mass of the compact
object and possibly reveal the presence of a spinning black hole.


\acknowledgements

We are grateful to Greg Sobczak for sharing with us his results prior
to publication. We also thank Mike Nowak, J\"orn Wilms, and Ramesh
Narayan for helpful comments on the manuscript. T.\,D.\,M.\
acknowledges support for this work provided by NASA through Chandra
Postdoctoral Fellowship grant number PF8-10005 awarded by the Chandra
Science Center, which is operated by the Smithsonian Astrophysical
Observatory for NASA under contract NAS8-39073. This work was
supported in part by a postdoctoral fellowship of the Smithsonian
Institution (D.\,P.).

\end{document}